\documentclass{article}
\usepackage[a4paper, margin=1in]{geometry} %
\usepackage{footnote}
\usepackage{rotating}
\usepackage{graphicx} 
\usepackage{color}
\usepackage{booktabs}
\usepackage{comment}
\usepackage{amsmath}
\usepackage{tabularx}
\usepackage{hyperref}
\usepackage{caption}
\usepackage{tablefootnote}
\makesavenoteenv{tabular}
\makesavenoteenv{table}
\usepackage{authblk}
\hypersetup{
	bookmarks=true,         
	unicode=false,          
	pdftoolbar=true,        
	pdfmenubar=true,        
	pdffitwindow=false,     
	pdfstartview={FitH},    
	pdftitle={My title},    
	pdfauthor={Author},     
	pdfsubject={Subject},   
	pdfcreator={Creator},   
	pdfproducer={Producer}, 
	pdfkeywords={keywords}, 
	pdfnewwindow=true,      
	colorlinks=true,       
	linkcolor=blue,          
	citecolor=blue,        
	filecolor=magenta,      
	urlcolor=cyan           
}

\newcommand{\tg}{\textcolor{black}}
\newcommand{\tr}{\textcolor{black}}
\newcommand{\tb}{\textcolor{black}}

\begin{document}

\title{Complete strategy spaces reveal hidden pathways to cooperation}

\author[1]{Zhao Song}

\author[1]{Ndidi Bianca Ogbo}

\author[2]{Xinyu Wang}

\author[3]{Chen Shen \thanks{ steven\_shen91@hotmail.com}}

\author[4,5,6,7]{Matjaz Perc \thanks{matjaz.perc@gmail.com}}

\author[1]{The Anh Han \thanks{T.Han@tees.ac.uk}}

\affil[1]{School of Computing, Engineering and Digital Technologies, Teesside University, United Kingdom}
\affil[2]{CSSC Systems Engineering Research Institute, China}
\affil[3]{Faculty of Engineering Sciences, Kyushu University, Japan}
\affil[4]{Faculty of Natural Sciences and Mathematics, University of Maribor, Maribor, Slovenia}
\affil[5]{Community Healthcare Center Dr. Adolf Drolc Maribor, Maribor, Slovenia}
\affil[6]{University College, Korea University, Seoul, Republic of Korea}
\affil[7]{Department of Physics, Kyung Hee University, Seoul, Republic of Korea}

\maketitle

\begin{abstract}
Understanding how cooperation emerges and persists is a central challenge in evolutionary game theory.
\tg{However, models of communication often rely on restricted strategy sets, making it unclear whether omitted behaviours alter cooperation.}
Here, we \tg{study} the complete deterministic eight-strategy space that naturally arises from the specified two-stage cheap-talk game.
We show, both analytically and through agent-based simulations, that incorporating the full deterministic strategy space fundamentally changes the evolutionary landscape.
\tg{In lattice populations, cooperation is sustained across wider parameter regions through previously overlooked cyclic dynamics involving suspicious cooperation, whereas strategic defection is fragile and mainly disrupts cooperation.}
\tg{The complete model also reveals stable coexistence among up to seven strategies and time-varying patterns of partial coexistence.}
These results demonstrate that the full strategy space unlocks hidden routes to cooperative behaviour, highlighting the importance of comprehensive modelling when explaining the emergence of cooperation.

\end{abstract}

\section*{Introduction}
The Evolutionary Game Theory (EGT) framework has provided the essential tools for understanding the emergence of cooperation among self-interested individuals~\cite{smith1982evolution,weibull1997evolutionary,hofbauer1998evolutionary}. Today, the persistence of cooperation in many strategic settings is better understood, supported by several well-established explanations   \cite{nowak2006five,perc2017statistical,HAN2025}, including  direct~\cite{nowak1993strategy,axelrod1981evolution} and indirect reciprocity~\cite{nowak1998evolution,santos2021complexity}, costly punishment~\cite{boyd2003evolution,hauert2007via}, and network reciprocity~\cite{nowak2006five,ohtsuki2006simple}. 
Built upon simplified yet ingenious strategies, these mechanisms offer crucial insights for various fields ranging from biology and social science to mathematics and computer science~\cite{apicella2019evolution,traulsen2023future,han2026socialphysicsageartificial}. This practice, however, comes at a cost: the choice of the strategy set can potentially conceal alternative pathways for cooperation to emerge~\cite{hauert2002volunteering,herrmann2008antisocial,rand2011evolution,garcia2019evolution}. This motivates a re-examination of a prior work on cheap talk that relies on a limited strategy set~\cite{song2025network}.

Cheap talk, defined as non-binding and cost-free communication prior to interaction without altering the underlying payoff structure~\cite{crawford1998survey,feldhaus2016more,andersson2012credible,gintis2001costly,nowak2006evolutionary,han2016emergence,duong2021cost}, presents a fundamental theoretical conundrum for the emergence of cooperation. 
\tr{Since signals are costless, standard game theory predicts rational individuals will deceive if profitable, rendering communication meaningless and defection prevailing. Paradoxically, this approach is ubiquitous and effective in human social life, ranging from verbally promising in team collaboration to expressing reliability in business activity. However, laboratory evidence consistently shows that while cheap talk can initially boost cooperation in one-shot interactions, this effect decays over time unless the interactions are repeated~\cite{arechar2017m,duffy2002actions,balliet2010communication}.} 

\tr{Consequently, it is crucial to understand the effect of cheap talk in one-shot scenarios without the influence of reputation.
Moreover, this is highly relevant today, as one-off interactions are increasingly common in the modern world, such as anonymous crowd-sourcing, ad-hoc gig economy tasks, and single-use digital transactions where long-term reputations are absent. To address this, we recently proposed a two-stage evolutionary game model as an insightful perspective, showing how cheap talk effectively sustains cooperation in one-shot scenarios by mediating local spatial interactions~\cite{song2025network}.}

Despite this success, our explanation of how cheap talk sustains cooperation is fundamentally constrained by its reliance on a limited four-strategy set. This simplification, while necessary for initial analysis, departs significantly from the complexity and potential variability found in real-world strategic systems~\cite{peysakhovich2017prosocial,hernandez2017survey}. 
\tr{More critically, a growing body of research argues that strategy selection is not a neutral modelling choice without consequence; direct comparisons have shown that restricted strategy spaces can artificially force specific evolutionary outcomes, often masking the true underlying pathways~\cite{garcia2025picking,sehwag2025collective,garcia2019evolution}. }This evidence compels us to ask: \textit{Does the simplified four-strategy model truly reveal the complete pathways for cooperation?}

To answer this question, we extend the prior work by developing a model of the complete \tb{deterministic} eight-strategy space \tb{of the specified two-stage cheap-talk game}, encompassing all possible actions derived from the two-stage game.
At the pre-game communication stage, players can choose whether to send a cooperative signal or remain silent, followed by choosing cooperation or defection at the in-game decision-making stage, which enables eight possible \tb{pure-strategy mappings ($2 \times 2 \times 2 = 8$)}.
\tb{Importantly, the prior four-strategy set was not a neutral truncation of this space: it excluded behaviourally admissible strategies that prove dynamically consequential.}
\tr{By combining analytical calculations of fixation probabilities and strategy abundances with agent-based simulations on structured populations  \cite{antal2009mutation,chen2024unbending,tarnita2009strategy,tarnita2011multiple,battiston2025higher}, we establish a framework in which \tb{weak-selection theory and simulations are compared at the level of qualitative trends, rather than quantitative abundance matching}.}
Critically, we demonstrate that the full strategy space is significantly more effective at sustaining cooperation than the simplified model. Furthermore, our findings fundamentally revise the understanding of the underlying pathways for cooperation. Specifically, we reveal a \tg{previously overlooked} cyclic dynamic involving the previously overlooked suspicious cooperation strategy, which catalyses cooperation, while the central biased component, strategic defection, is found to be competitively fragile and acts primarily as a spoiler. Finally, we uncover richer evolutionary pathways, including the stable coexistence of up to seven distinct strategies and alternating patterns of coexistence over time.
Our work thus underscores the necessity of fully exploring the strategy space to accurately map the underlying conditions for cooperation.

\section*{Results}
Here, we first test the role of cheap talk on sustaining cooperation in a finite well-mixed population without network reciprocity mechanisms. Then we examine the success of cheap talk when combined with network reciprocity on a lattice network.
\subsection*{Well-mixed population}
\begin{figure}
    \centering
    \includegraphics[width=\textwidth]{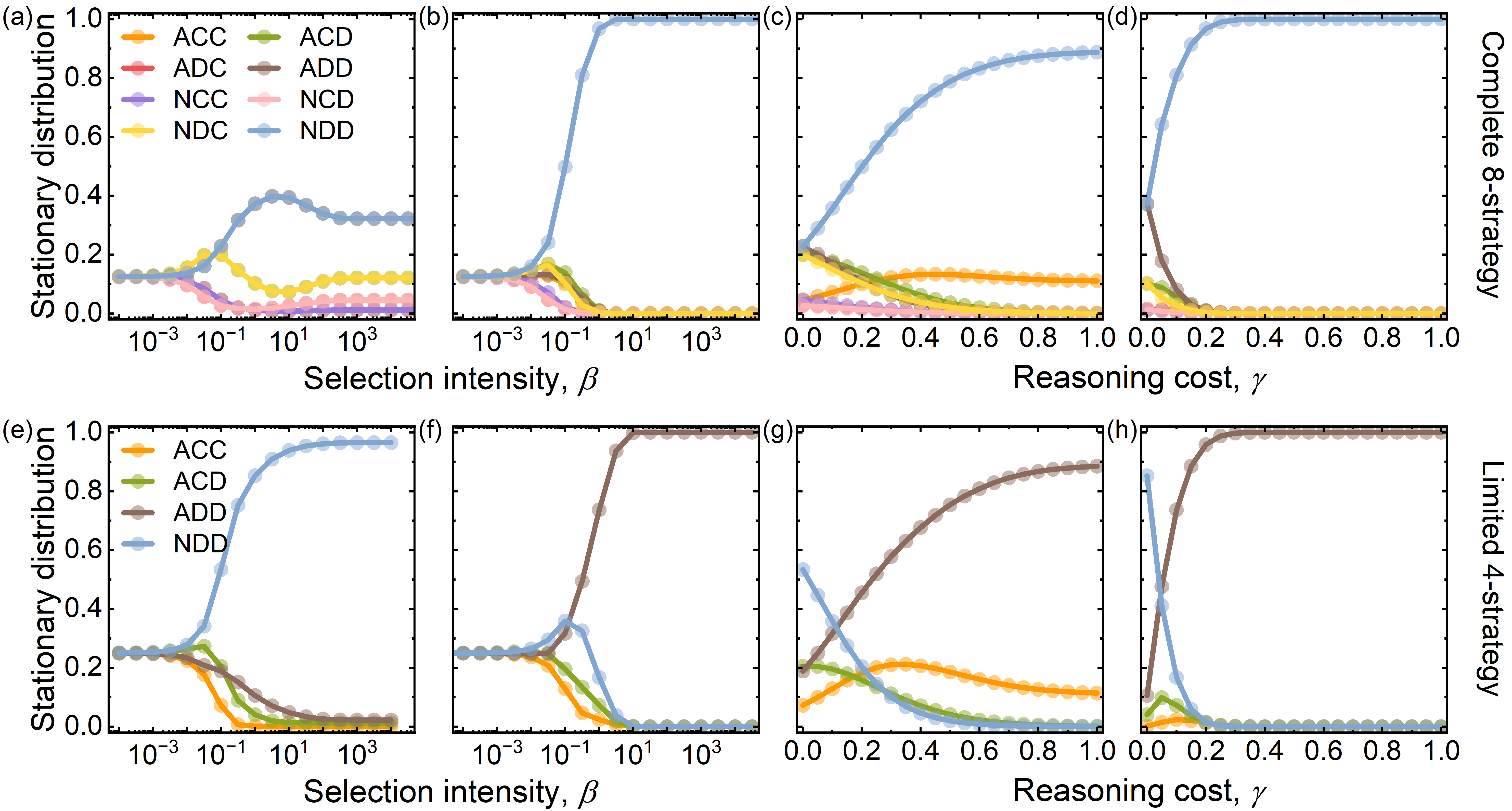}
    \caption{}
    \label{fig1}
\end{figure}

In the finite well-mixed population without any reciprocity mechanism, cheap talk shows limited effectiveness in sustaining cooperation. Across a wide range of selection intensities and reasoning costs, defective strategies dominate. When reasoning cost is absent ($\gamma=0$, Figure \ref{fig1}\tg{a}), at weak selection intensities ($\beta<10^{-2}$), eight strategies co-exist equally due to the neutral transitions. However, as the selection intensity strengthens, strategic defection and unconditional defection (ADD and NDD) quickly take over the population. A similar pattern holds when a reasoning cost is introduced ($\gamma=0.2$, Figure \ref{fig1}\tg{b}); eight strategies are equally distributed with weak selection intensity, while NDD prevail as soon as the selection intensity becomes strong (around $\beta>10^{-3}$). Under weak selection ($\beta=0.1$, Figure \ref{fig1}\tg{c}), some cooperative strategies (for instance, ACC and ACD) can co-exist with defective strategies when the reasoning cost is small (around $\gamma<0.7$). As the reasoning cost grows, the system simplifies to a mix of intuitive cooperation and defection (ACC and NDD). This small window for cooperation vanishes completely under strong selection ($\beta=1$, Figure \ref{fig1}\tg{d}), where NDD dominates as soon as the cost of reasoning eliminates deliberative strategies.

\tr{
Compared to the limited 4-strategy model, though defective strategy dominates, cooperation survives through strategies ACD and NDC when the reasoning cost is 0 (Figure \ref{fig1}\tg{e}). On the other hand, ADD, the dominant strategy in the limited 4-strategy model, becomes fragile in the complete 8-strategy model; instead, NDD becomes the dominant defective strategy (Figure \ref{fig1}\tg{f-h}).
In addition, as detailed in \tg{Supplementary Note 2}, we confirm that in the well-mixed population, NDD universally has the highest abundance in the parameter space in the complete 8-strategy model.}

\subsection*{Lattice population}
\begin{figure}
    \centering
{\includegraphics[width=0.9\textwidth]{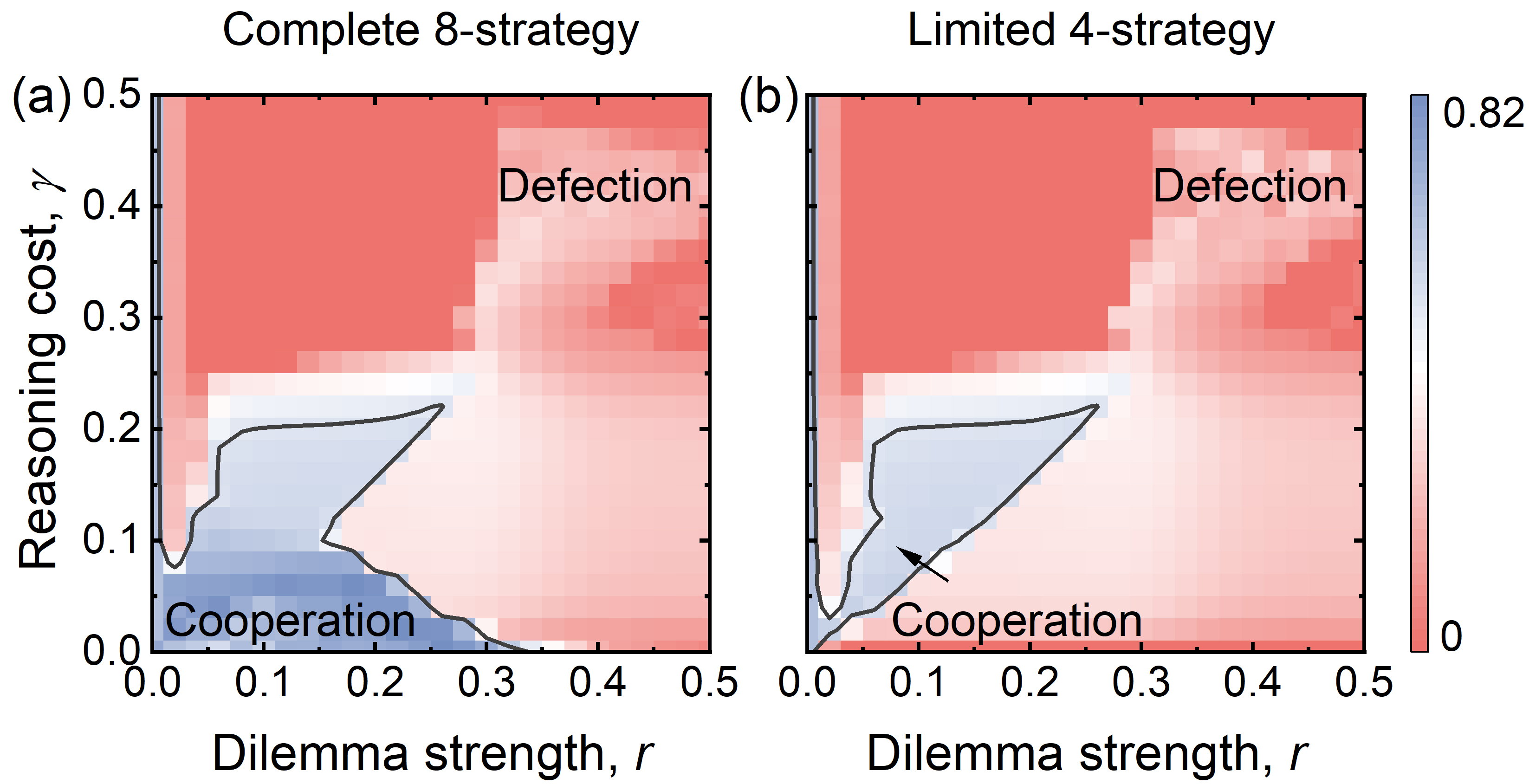}} 
\caption{}
    \label{fig2}
\end{figure}
In stark contrast to the well-mixed population, introducing network reciprocity allows cheap talk to become a powerful force for cooperation, especially when the full space of strategies is available. This effect is particularly prominent when the dilemma is moderately difficult and reasoning costs are low. In the full strategy space (Figure \ref{fig2}\tg{a}), a robust region of high cooperation (blue area) emerges, spanning from dilemma strength $r=0$ to approximately $r=0.3$. This cooperative phase is sustained for reasoning costs up to $\gamma=0.25$. In contrast, the simulation with the limited strategy space (Figure \ref{fig2}\tg{b}) shows a much smaller and more fragile region of cooperation. Here, high cooperation is confined to lower dilemma strengths, diminishing rapidly after $r>0.15$, and collapses at a similar reasoning cost threshold. In both scenarios, cooperation gives way to widespread defection (red area) as either the dilemma strength or the reasoning cost becomes too high.

\subsubsection*{Weak dilemma strength}
\begin{figure}[tb]
    \centering
{\includegraphics[width=\textwidth]{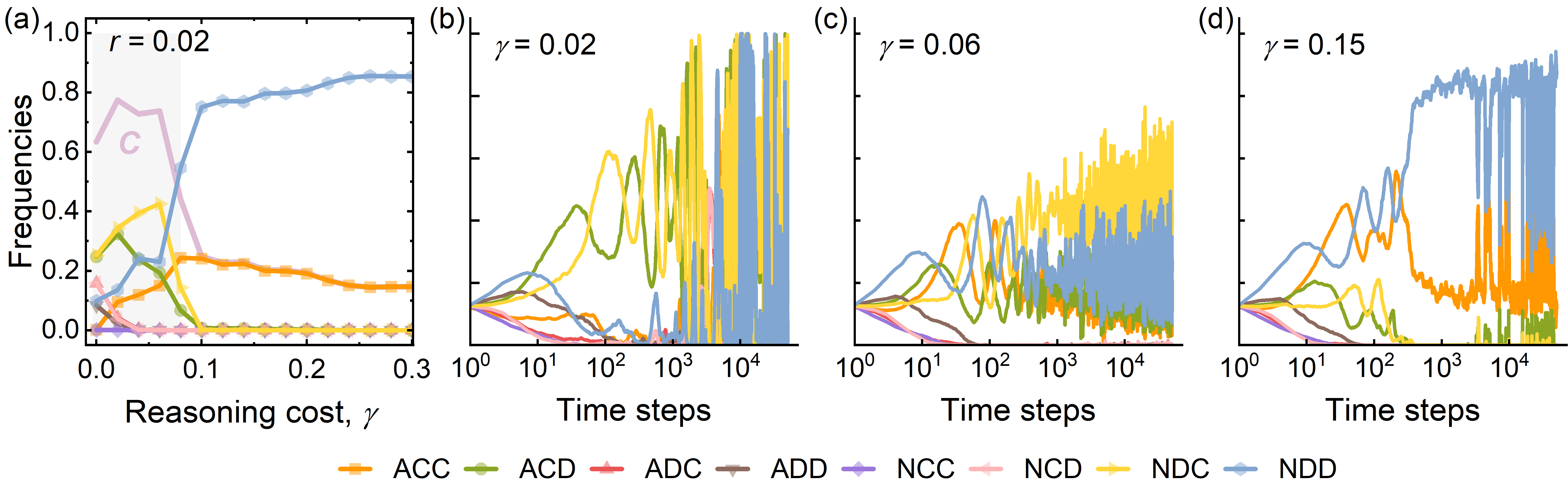}} 
\caption{}
    \label{fig2-1}
\end{figure}

\begin{figure}[tb]
    \centering
    \includegraphics[width=\linewidth]{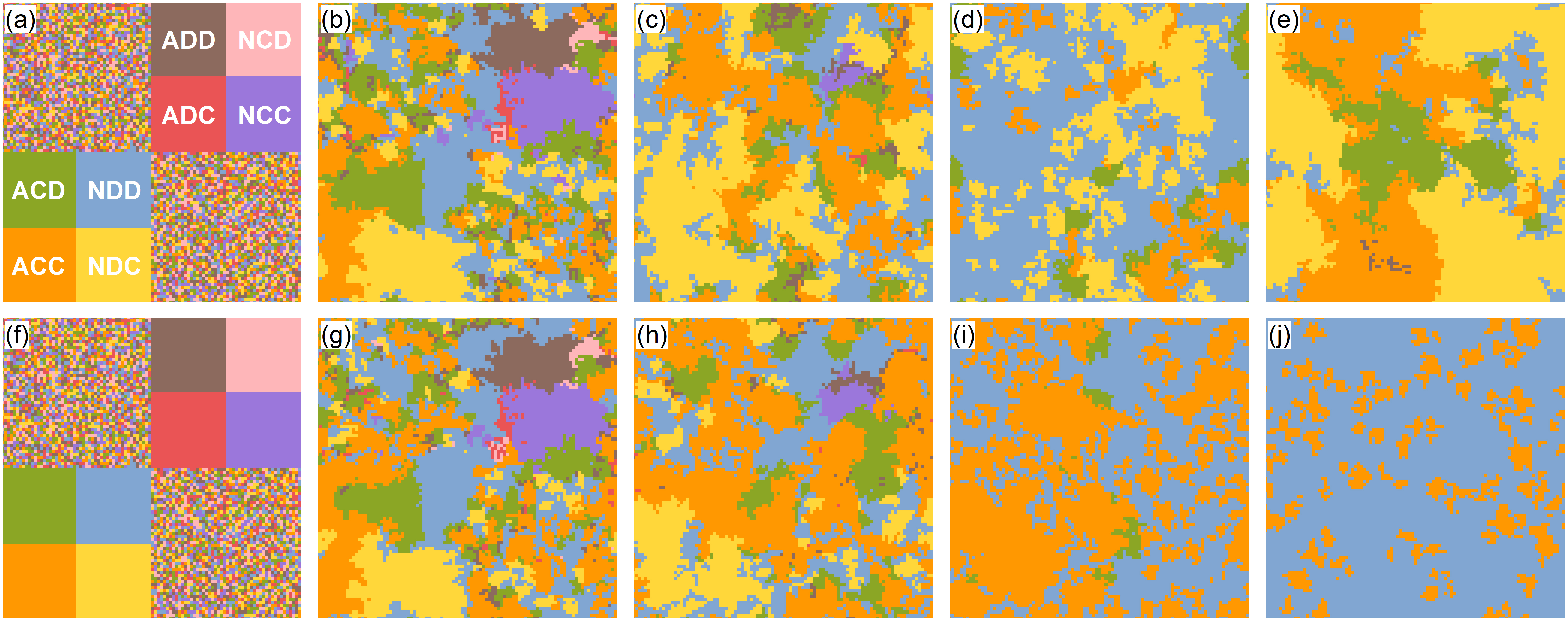}
    \caption{}
    \label{fig5-1}
\end{figure}

In a weak dilemma ($r=0.02$), cheap talk's ability to sustain cooperation is fundamentally dependent on the reasoning cost. When reasoning is inexpensive, cheap talk successfully promotes cooperation by enabling a rich ecosystem of conditional strategies (Figure \ref{fig2-1}\tg{a}). \tr{To compare, the results of the limited 4-strategy model can be found in \tg{Supplementary Figure S1a in Supplementary Note 1.}} At low reasoning cost, cooperation is sustained by the complex coexistence of seven strategies except NCC ($\gamma=0.02$, Figure \ref{fig2-1}\tg{b}). As the cost increases, this ecosystem simplifies, where four ACC, ACD, NDC, and NDD co-exist, but cooperation is still upheld by strategies reliant on cheap talk ($\gamma=0.06$, Figure \ref{fig2-1}\tg{c}). However, when the reasoning cost becomes too high, the mechanism of cheap talk completely breaks down ($\gamma=0.15$, Figure \ref{fig2-1}\tg{d}). The cost of strategic deliberation purges all conditional strategies, rendering communication ineffective. In this regime, the population simplifies to only unconditional cooperators (ACC) and defectors (NDD). Cooperation persists, but it is sustained solely by the fundamental mechanism of network reciprocity. 

The spatial evolution of strategies reveals that cheap talk sustains cooperation by the cyclic dominance of four strategies, involving unconditional cooperation, unconditional defection, conditional cooperation, and suspicious cooperation\footnote{See detailed dynamics at: \url{https://osf.io/r6je4/?view_only=4396b17faba74b87b9454972e243a164}}. When the reasoning cost is moderate $\gamma=0.06$ (Figure \ref{fig5-1}\tg{a-e}), from a mixed initial state, clusters of strategies, including ACC, ACD, NDC, and NDD, form and compete. The persistence of cooperation is driven by a form of cyclic dominance: NDD invades NDC clusters, but these victorious NDD clusters are in turn invaded by the conditional cooperator ACD. The ACD clusters are then replaced by ACC, which are finally invaded by NDC, completing the cycle. This rock-paper-scissors dynamic prevents any single strategy from dominating and, crucially, allows cooperative strategies to continually reclaim space from defectors. However, this complex dynamic collapses when the reasoning cost is too high for deliberative strategies to survive ($\gamma=0.15$, Figure \ref{fig5-1}\tg{f-j}). The deliberative strategies that drove the cycle are quickly eliminated. The population simplifies to only two unconditional strategies, ACC and NDD. Here, cooperation persists through a simpler mechanism: the benefits of network reciprocity allow ACC players to form stable, segregated clusters that can resist invasion from the surrounding NDD players. 
\tr{Additionally, in \tg{Supplementary Note 3}, we analytically demonstrate that the full \tb{deterministic} strategy space significantly enlarges the parameter regions where cooperative strategies (particularly ACD and the \tg{the previously omitted} NDC) achieve the highest abundance.}

\subsubsection*{Strong dilemma strength}
\begin{figure}[tb]
    \centering
{\includegraphics[width=\textwidth]{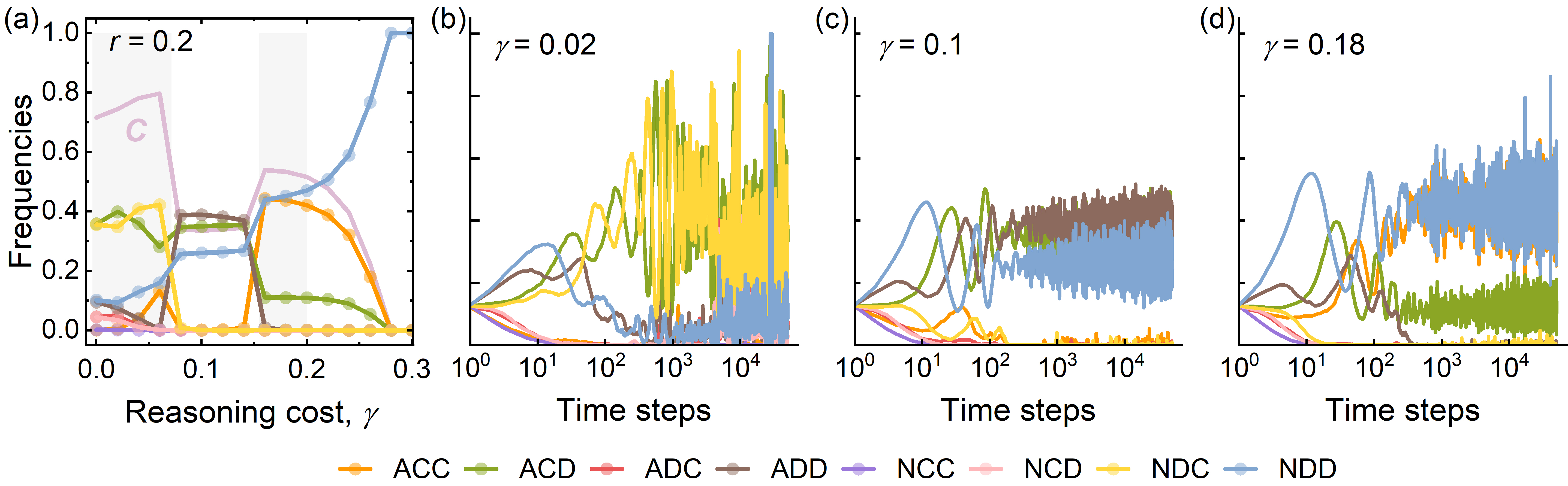}} 
\caption{}
    \label{fig2-3}
\end{figure}

\begin{figure}[tb]
    \centering
    \includegraphics[width=\linewidth]{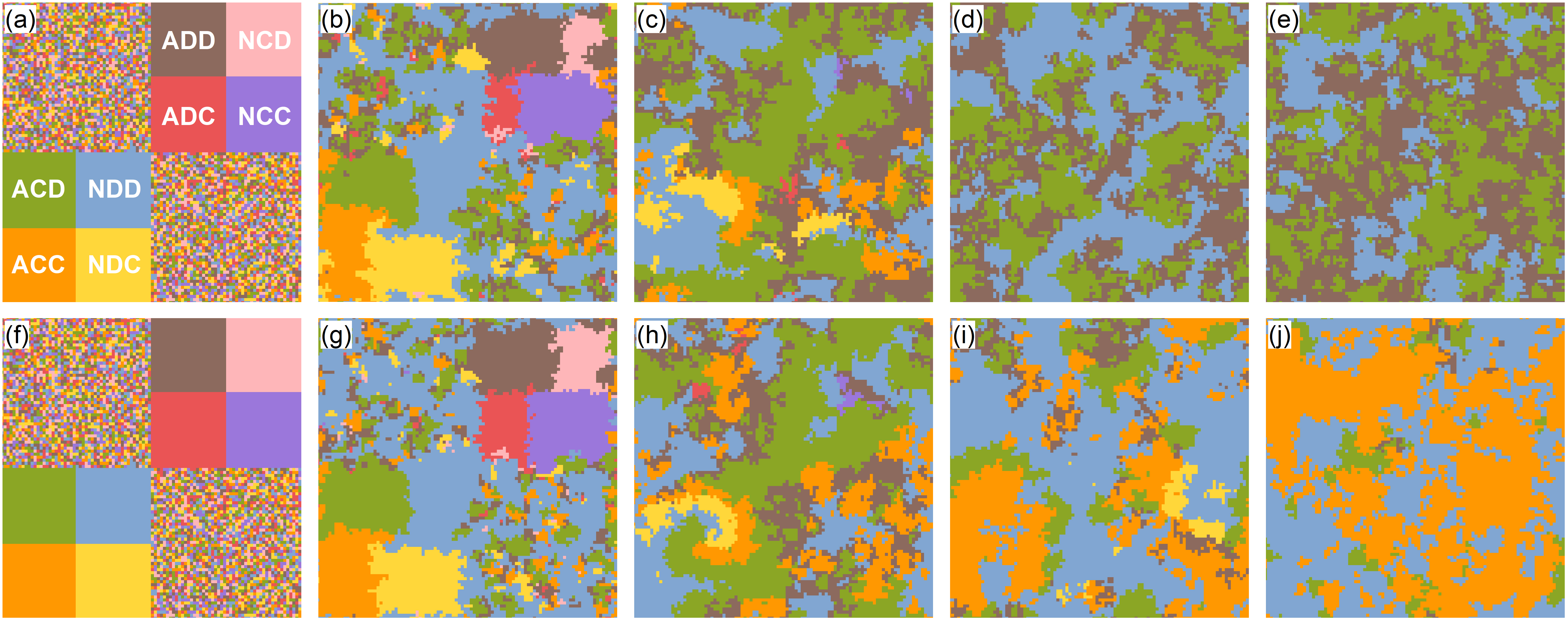}
    \caption{}
    \label{fig5-2}
\end{figure}

When the dilemma is strong ($r=0.2$), cooperation is sustained via complex co-existence of strategies but will be undermined by strategic defection. \tr{To compare, the results of the limited 4-strategy model can be found in \tg{Supplementary Figure S1b in Supplementary Note 1.}} For intermediate reasoning costs (around $0.08<\gamma<0.2$), the strategic defection ADD thrives, creating a deep ``valley" of low cooperation (Figure \ref{fig2-3}\tg{a}). By falsely signalling cooperative intent, ADD players corrupt the meaning of communication and exploit cooperative opponents, as seen in their rise to dominance in the evolutionary trajectory (Figure \ref{fig2-3}\tg{c}). This deceptive regime only exists in a specific reasoning cost window. At lower costs ($\gamma=0.02$, Figure \ref{fig2-3}\tg{b}), the co-existence of seven strategies can sustain a higher level of cooperation. While at higher costs  ($\gamma=0.18$, Figure \ref{fig2-3}\tg{d}), cooperation dominates via the cyclic dominance among ACC, ACD, and NDD. Furthermore, in contrast to the weak dilemma scenario, cooperation completely collapses for high reasoning costs ($\gamma>0.25$). With a strong dilemma strength, network reciprocity is no longer sufficient to prevent the extinction of ACC by NDD.

\begin{figure}[tb]
    \centering
    \includegraphics[width=\linewidth]{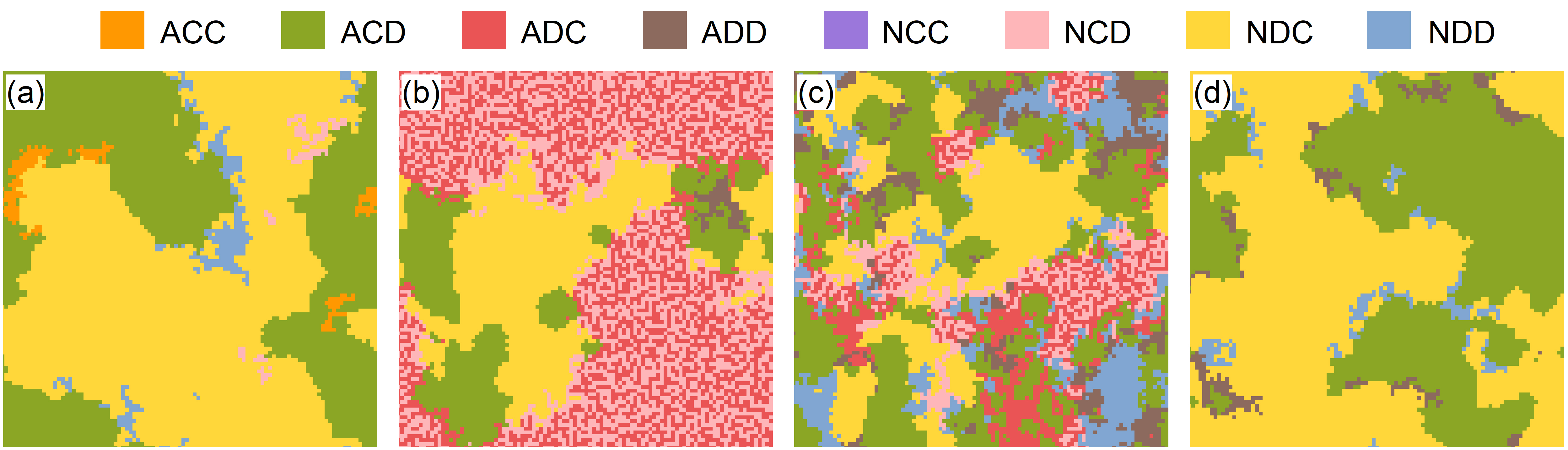}
    \caption{}
    \label{fig6-1}
\end{figure}

The spatial dynamics under a strong dilemma reveal that in strong dilemmas, cooperation is undermined by a cyclic dynamic involving strategic defection strategies. In detail, in the ``valley of deception" ($\gamma=0.1$, Figure \ref{fig5-2}\tg{a-e}), cooperation is actively suppressed by a rock-paper-scissors dynamic. The cycle unfolds as follows: conditional cooperators ACD, can invade clusters of unconditional defectors NDD. However, ACD's strategy of trusting the `$A$' signal makes it a perfect target for the strategic ADD strategy, which systematically exploits and invades it. Completing the cycle, the cost-free NDD players have a direct fitness advantage over the cognitively costly ADD players, allowing them to invade in turn. This constant cycle prevents any single strategy from dominating and creates a landscape where cooperation is undermined since it can only survive within ACD clusters. However, if the reasoning cost is higher ($\gamma=0.18$, Figure \ref{fig5-2}\tg{f-j}), the strategic ADD strategy is too costly to survive, and a different, more pro-cooperative cycle emerges. Here, ACD acts as the catalyst that continually disrupts the large NDD clusters. This creates the necessary space for unconditional cooperators ACC to form their own stable clusters, leading to a dynamic state where cooperation can persist within both ACC and ACD clusters. 
Furthermore, when the reasoning cost is small ($\gamma=0.02$, Figure \ref{fig6-1}\tg{a},\tg{b}), cooperation is sustained via dynamical co-existence of strategies. Surprisingly, ADC and NCD co-exist, interacting with each other using cooperation.
In addition, when the reasoning cost is small ($\gamma=0.02$, Figure \ref{fig6-1}\tg{c,d}), cooperation is sustained via complex and various strategy co-existence patterns.

\begin{figure}
    \centering
    \includegraphics[width=0.8\linewidth]{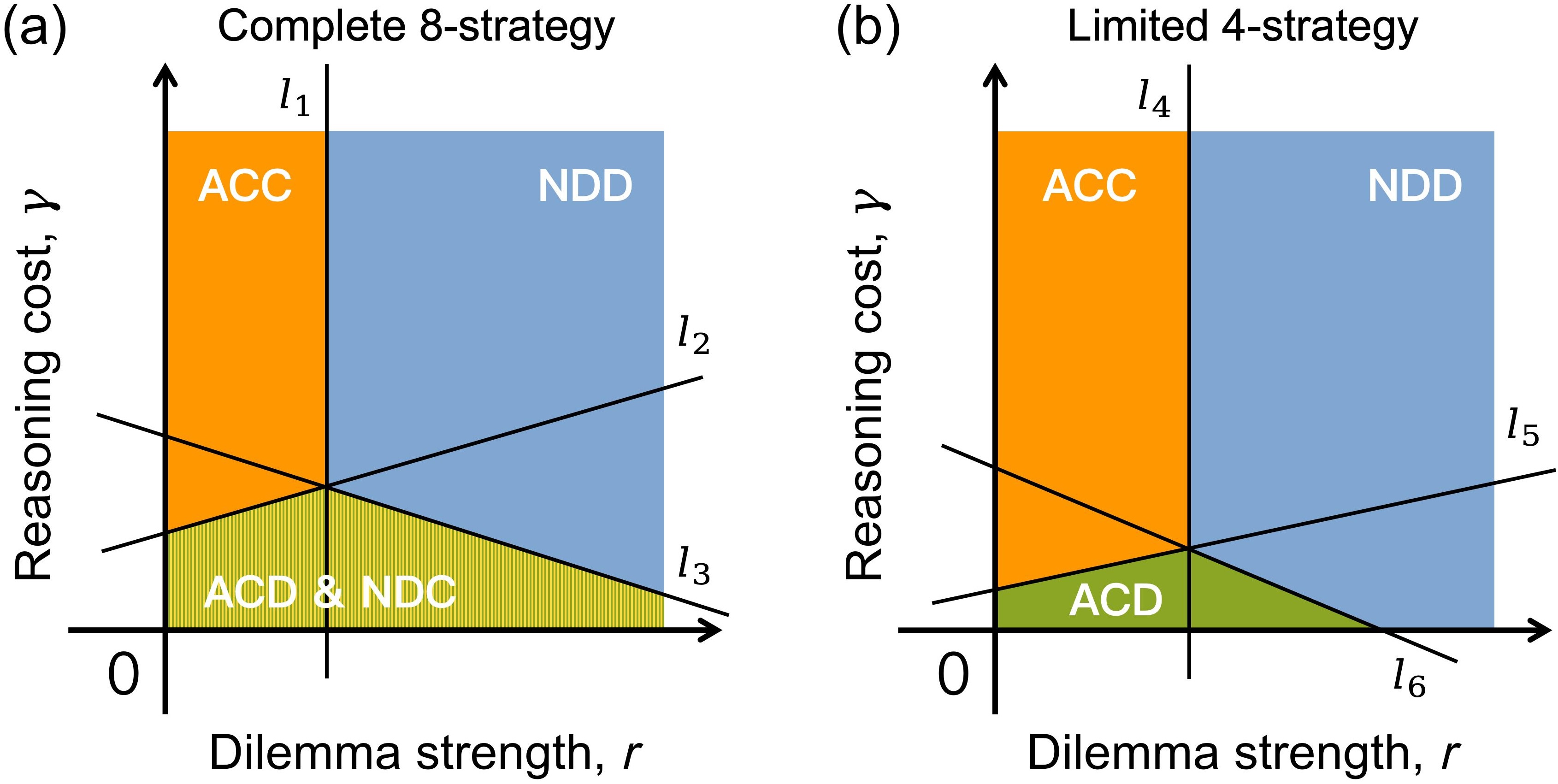}
    \caption{}
    \caption{\tr{\textbf{\tg{Additional} pathways for cooperation emerge in the complete model.} The parameter regions indicating the strategy with the highest abundance in a square lattice population,  for (a) the complete 8-strategy model and (b) the limited 4-strategy model. The solid lines indicate, $l_1:\gamma=\frac{1}{3}$, $l_2:\gamma=\frac{6r+3}{16}$, $l_3:\gamma=\frac{-6r+7}{16}$, $l_4:\gamma=\frac{8}{21}$, $l_5:\gamma=\frac{6r+3}{32}$, $l_6:\gamma=\frac{-15r+11}{32}$. The complete 8-strategy model contains strategies ACC, ACD, ADC, ADD, NCC, NCD, NDC, and NDD; the limited 4-strategy model contains strategies ACC, ACD, ADD, and NDD.}}
    \label{figA2}
\end{figure}

\tr{
\subsubsection*{Strategy selection}
To conclude, we examine the evolutionary dynamics under the limit of weak selection, offering theoretical insights based on strategy abundance \cite{antal2009mutation,tarnita2009strategy,tarnita2011multiple}. 
\tb{These analytical predictions are derived in the weak-selection limit ($\beta\to 0$) and provide qualitative guidance rather than quantitative fits to our simulations, which use $\beta=10$.} 
As illustrated in Figure \ref{figA2}, the NDC strategy—which is absent in the limited model—emerges as a \tg{previously overlooked} pathway for cooperation within the complete strategy model. Although the absolute parameter space dominated by the pure cooperator ACC shrinks slightly in the complete model relative to the limited one, the overall prevalence of cooperation is substantially enhanced. This improvement is primarily driven by ACD and, crucially, NDC, which together occupy a substantial fraction of the parameter space. \tb{Because the first-order weak-selection criterion treats ACD and NDC identically, the distinctive catalytic role of NDC is identified through the spatial simulations above. These analytical results are consistent with the qualitative trends observed in our lattice simulations:} the complete strategy model reveals the pathways for cooperation, while the previous limited models failed to capture.
Detailed analyses for both well-mixed and lattice populations are included in \tg{Supplementary Notes 2 and 3.}
}

\section*{Discussion}

In this study, we extended a prior four-strategy model of cheap talk to the complete \tb{deterministic} eight-strategy space \tb{of the specified two-stage game} in order to uncover the full evolutionary pathways that sustain cooperative behaviour. Our analysis began by confirming a baseline result: in well-mixed populations without any reciprocity mechanism, cheap talk remains largely ineffective, as defective strategies rapidly dominate when selection intensity or reasoning cost increases. In contrast, in structured populations where network reciprocity operates, we found that the full strategy space is markedly more effective at supporting cooperation than the restricted four-strategy model. This enhanced performance arises from qualitatively \tg{previously overlooked} pathways, including a cooperative cycle in weak dilemmas driven by the previously overlooked suspicious cooperation strategy; the fragility of strategic defection, which acts primarily as a spoiler in strong dilemmas; and complex coexistence patterns that persist at very low reasoning costs.

Extending the model to the complete \tb{deterministic} strategy space substantially advances our understanding of how cheap talk interacts with network reciprocity to sustain cooperation. While the simplified model correctly identified network reciprocity as a prerequisite for cheap talk to exert influence, its constrained design led both to a quantitative underestimation of cheap talk's stabilising power and to a qualitative mischaracterisation of the mechanisms involved. The complete strategy set reveals that the true catalyst for cooperation is a cyclic dynamic driven by suspicious cooperation -- a counterintuitive strategy that never signals, defects upon receiving a signal, yet cooperates when encountering silence. This behaviour, entirely absent from the simplified framework, fundamentally reshapes the cooperative routes available to the population. Likewise, the full model clarifies the nuanced role of strategic defection, a strategy that signals but never cooperates: rather than forming a robust component of the evolutionary ecosystem, it is competitively fragile and primarily disruptive to cooperative outcomes.

Our work thus validates the methodological necessity of adopting a complete \tb{deterministic} strategy space by adhering to the core principles for strategy selection: ensuring computational equivalence and the absence of bias across strategies; constructing a comprehensive microeconomic model of interactions; and grounding the model in stylised facts~\cite{garcia2025picking}. First, we include all eight strategies that can arise from the two-stage game, avoiding the arbitrary exclusion of non-abundant or counterintuitive behaviours. Second, our framework incorporates a full cheap-talk mechanism in which each player can send a non-binding signal and condition their action on the co-player's signal or silence. Third, our model is informed by the stylised fact that cheap talk can support cooperation in real social settings, and we identify the evolutionary conditions under which this effect emerges when combined with network reciprocity. These results demonstrate that strategy selection is not a neutral modelling choice: the inclusion or omission of specific strategies materially affects the ecological dynamics observed, from cyclic competition and contingent cooperation to the emergence of deception.

More broadly, both limited and complete strategy sets contribute to the ongoing effort to understand how cooperative behaviour arises, yet our findings underscore a dual lesson for evolutionary modelling: transparency regarding strategy inclusion is essential, and the potential relevance of seemingly peripheral or overlooked strategies warrants continual re-evaluation through evolutionary analysis.

Finally, although our eight-strategy model provides an unbiased representation of the \tb{deterministic pure-strategy space of the} two-stage cheap-talk game, the notion of completeness remains relative in the broader context of realistic social interactions~\cite{tkadlec2023mutation}. Here, we assumed perfect, noise-free communication, whereas real interactions are often subject to misinterpretation and stochastic errors. \tb{Therefore, beyond the scope of our model, stochastic strategies, noisy signals, multi-signal communication, and richer behavioural rules represent important extensions.} 
Moreover, our model considers only two possible signals and two possible response types, whereas real-world communication is vastly richer. Extending the framework to multi-signal and multi-action spaces may reveal more elaborate cheap-talk protocols and uncover \tg{additional} mechanisms for shaping cooperative outcomes~\cite{santos2011co}.

\section*{\tg{Methods}}
\subsection*{Two-stage game for cheap talk}
We model the interaction with cheap talk as a two-stage game. The first stage is the pre-game communication, where players can either announce an intention to cooperate ($A$) or remain silent ($N$). This signal is non-binding and carries no intrinsic cost. Following the communication, players enter the in-game decision-making stage, which is structured as a Donation Game, a classic social dilemma. In this stage, players simultaneously choose to either cooperate ($C$) or defect ($D$). A player who cooperates incurs a cost $c > 0$ to provide a benefit $b$ to their co-player, where $b>c$. A defector pays no cost and provides no benefit. This leads to four standard payoffs: mutual cooperation yields a reward $R=b-c$; mutual defection results in a punishment $P=0$; and a defector exploiting a cooperator gets the temptation payoff $T=b$, leaving the cooperator with the sucker's payoff $S=-c$. By normalising these values, we can express the game in the form of a Prisoner's Dilemma with $R=1$, $P=0$, $S=-r$, and $T=1+r$. \tr{The parameter $r=c/(b-c)>0$  quantifies the strength of the dilemma}~\cite{wang2015universal}.

A player's complete strategy can be represented as a triplet defining their actions in all possible scenarios: i) their initial signal ($A$ or $N$); ii) their action ($C$ or $D$) if the opponent signals $A$; and iii) their action ($C$ or $D$) if the opponent signals $N$. This formulation gives rise to eight distinct \tb{pure} strategies, as detailed in Table \ref{table:strategy}. \tb{Throughout the paper, `complete strategy space' refers to the complete deterministic pure-strategy space of the specified binary, noise-free, two-stage cheap-talk game, not to the unrestricted set of all stochastic, mixed, noisy, memory-dependent, or multi-signal communication strategies.}

\begin{table}[tb]
\centering
\caption{\tg{Eight strategies in the two-stage game with cheap talk.}}
\begin{tabularx}{\textwidth}{lXXX}
\toprule
Strategy & Signal cooperative intention? & Cooperate if the co-player signals? & Cooperate if the co-player remains silent? \\
\midrule
ACC & Yes & Yes & Yes \\
ACD & Yes & Yes & No  \\
ADC & Yes & No  & Yes \\
ADD & Yes & No  & No  \\
NCC & No  & Yes & Yes \\
NCD & No  & Yes & No  \\
NDC & No  & No  & Yes \\
NDD & No  & No  & No  \\
\bottomrule
\end{tabularx}
\label{table:strategy}
\end{table}

These strategies can be categorised into two types: intuitive strategies, including ACC (always signal and cooperate unconditionally) and NDD (never signal and always defect);  and deliberative strategies, including the remaining strategies, making their actions contingent on the opponent's signal. We model the cognitive effort of such conditional logic by imposing a reasoning cost, $\gamma$, on these deliberative strategies, where $0\leq\gamma\leq1$. This cost ensures that strategic thinking is not without its trade-offs. The model's dynamics are sensitive to this cost: if $\gamma=0$, deliberative strategies can leverage their flexibility without penalty. Conversely, if the reasoning cost is high, these costly strategies become unviable, and the game simplifies to an interaction between unconditional cooperators and defectors. The comprehensive payoff matrix for interactions between all eight strategies is given in Equation (\ref{payoffMatrix}):
\begin{equation}
    \begin{tabular}{c|cccccccc}
           & ACC& ACD & ADC &ADD  & NCC& NCD & NDC & NDD\\
           \hline
        ACC & $R$ & $R$ &$S$ &$S$ &$R$ &$R$ &$S$ &$S$\\
        ACD &$R-\gamma$ & $R-\gamma$ &$S-\gamma$ &$S-\gamma$ &$T-\gamma$ &$T-\gamma$ &$P-\gamma$ &$P-\gamma$\\
        ADC & $T-\gamma$ & $T-\gamma$ &$P-\gamma$ &$P-\gamma$ & $R-\gamma$ & $R-\gamma$ &$S-\gamma$ &$S-\gamma$ \\
        ADD & $T-\gamma$ & $T-\gamma$ &$P-\gamma$ &$P-\gamma$ &$T-\gamma$ &$T-\gamma$ &$P-\gamma$ &$P-\gamma$\\
        NCC & $R-\gamma$ & $S-\gamma$ &$R-\gamma$ &$S-\gamma$ & $R-\gamma$ & $S-\gamma$ &$R-\gamma$ &$S-\gamma$\\
        NCD & $R-\gamma$ & $S-\gamma$ &$R-\gamma$ &$S-\gamma$ & $T-\gamma$ & $P-\gamma$ &$T-\gamma$ &$P-\gamma$\\
        NDC & $T-\gamma$ & $P-\gamma$ &$T-\gamma$ &$P-\gamma$ & $R-\gamma$ & $S-\gamma$ &$R-\gamma$ &$S-\gamma$\\
        NDD & $T$ & $P$ &$T$ &$P$ & $T$ & $P$ &$T$ &$P$
    \end{tabular}.
    \label{payoffMatrix}
\end{equation}

\subsection*{Well-mixed finite population}
We first analyse the evolutionary dynamics within a finite, well-mixed population of $M$ individuals. In this scenario, every player is equally likely to interact with any other player. Strategy updates are governed by the Moran process. In each time step, one individual is randomly selected to update its strategy, and a second individual is randomly chosen as a potential role model.

Assuming the population consists of players adopting strategies $A$ and $B$, with $m$ players adopting strategy $A$ and $M-m$ players adopting strategy $B$. The average payoffs $f_A$ and $f_B$ for an individual with strategy $A$ and strategy $B$ are:
\begin{equation}
\begin{split}
     & f_{A}=\frac{(m-1)\pi_{A,A}+(M-m)\pi_{A,B}}{M-1}, \\
     & f_{B}=\frac{m\pi_{B,A}+(M-m-1)\pi_{B,B}}{M-1},
\end{split}
\end{equation}
where $\pi_{X,Y}$ denotes the payoff when the focal player adopting strategy $X$ interacts with a co-player adopting strategy $Y$.

The probability that the player with strategy $A$ imitates the strategy of the player with strategy $B$ is determined by the Fermi function, $(1+e^{\beta (f_{A}-f_{B})})^{-1}$~\cite{traulsen2007pairwise}. The parameter $\beta$ is the selection intensity, which controls how strongly payoff differences influence imitation. When $\beta\rightarrow \infty$, players deterministically copy the strategy with the higher payoff, while for $\beta=0$, imitation is random (neutral drift).

\tr{The probability that the number $m$ of strategy $A$ in the population increases or decreases by one are given by $T^{+}(m)$ and $T^{-}(m)$, respectively:
\begin{equation}
       T^{\pm}(m)=\frac{M-m}{M}\frac{m}{M}[1+e^{\mp \beta (f_{A}-f_{B})}]^{-1}.
\end{equation}
}
Using these transition probabilities, we can calculate the fixation probability, $\rho_{BA}$, which is the probability that a single mutant with strategy $A$ will eventually take over a resident population of $M-1$ players with strategy $B$\cite{sigmund2010social}:
\tr{
\begin{equation}
    \rho_{BA}=\frac{1}{1+\sum_{k=1}^{M-1}\prod_{m=1}^{k}\frac{T^-(m)}{T^+(m)}}.
\end{equation}
}
Under the assumption of a low mutation rate, the evolutionary process can be modelled as a Markov chain on the set of homogeneous population states\cite{nowak2004emergence,imhof2005evolutionary}. The transition probabilities between these states are derived from the fixation probabilities. The stationary distribution of this Markov chain, found by calculating the normalised eigenvector of the transposed transition matrix, reveals the long-term proportion of time the population spends in each of the homogeneous states. \tr{The transition matrix is denoted as $\Lambda_{AB,A\neq B}=\rho_{AB}/(q-1)$ and $\Lambda_{AA}=1-\sum_{B=1, B\neq A}^q\Lambda_{AB}$, where $q$ is the number of strategies.}

\subsection*{Lattice finite population}
To investigate the role of network reciprocity, we also model the dynamics on the $L\times L$ square lattice with periodic boundary conditions. In this configuration, each player occupies a node on the lattice and interacts only with their four immediate neighbours (the von Neumann neighbourhood). The simulation proceeds via an asynchronous Monte Carlo algorithm. Initially, players are assigned one of the eight strategies with equal probability. In each generation, every player accumulates a total payoff, $\phi$, from interacting with their four neighbours.

The strategy update protocol is as follows: in each elementary step, a focal player $i$ is chosen at random. With a small probability $\mu$, the player undergoes mutation and adopts a randomly selected strategy. With probability $1-\mu$, player $i$ attempts imitation. For this, a random neighbour $j$ is selected, and player $i$ adopts player $j$'s strategy with a probability given by the Fermi function:
\begin{equation}
    F(j\rightarrow i)=\frac{1}{1+e^{\beta(\phi_i-\phi_j)}},
    \label{eq:fermi}
\end{equation}
where $\phi_i$ and $\phi_j$ are the total payoffs of player $i$ and $j$, respectively. One Monte Carlo step (MCS) consists of $L^2$ such elementary updates, meaning each player is chosen to update their strategy once on average.

For our simulations, we use a population size of $L^2=200^2$. The selection intensity is set to $\beta=10$ to model strong selection. 
\tr{A low mutation rate of $\mu=10^{-5}$ is included to mitigate the finite-size effects ~\cite{shen2025mutation}.} Simulations are run for $3\times 10^4$ MCS, and data are averaged over the final 5,000 steps to ensure the system has reached an evolutionary equilibrium.
\tb{The analytical results in \tg{Supplementary Notes 2 and 3} are derived in the weak-selection limit. We compare the two approaches at the level of qualitative trends (e.g., which strategies dominate which parameter regions), not quantitative abundance matching.} 

\section*{\tg{Acknowledgements}}
\tg{The authors have no additional acknowledgements.}

\section*{\tg{Funding}}
This work was supported by EPSRC (grant EP/Y00857X/1) to Z.S. and T.A.H., and JSPS KAKENHI (Grant no. JP 23H03499) to C.\,S.. M.P. was supported by the Slovenian Research and Innovation Agency (Grant No. P1-0403).

\section*{\tg{Author contributions}}
\tg{Z.S., C.S. and T.A.H. conceived and designed the study. Z.S. and X.W. performed the simulations and analyses. Z.S., N.B.O., X.W., C.S., M.P. and T.A.H. interpreted the results. Z.S. wrote the initial draft, and all authors reviewed and edited the manuscript.}

\section*{\tg{Competing interests}} \tg{The authors declare no competing interests.}

\section*{\tg{Data availability}}
\tg{The data generated and analysed in this study are available from the corresponding authors upon request.}

\section*{\tg{Figure captions}}

\noindent\textbf{Figure 1. Cheap talk alone can not sustain cooperation in finite well-mixed populations.} \tr{\tg{Panels a and b (e and f)} show the stationary distribution of each strategy as a function of selection intensity; \tg{panels c and d (g and h) } show the stationary distribution of each strategy as a function of reasoning cost, each \tb{at a fixed selection intensity}. \tg{Panels a-d show the results of the complete 8-strategy model and panels e-h show the results of the limited 4-strategy model.} Results are obtained with parameters $r=0.2$, \tg{a and e}: $\gamma=0$; \tg{b and f}: $\gamma=0.2$; \tg{c and g}: $\beta=0.1$, and \tg{d and h}: $\beta=1$.}

\noindent\textbf{Figure 2. A full strategy space makes cheap talk more effective at promoting network cooperation.} Panel \tg{a} shows the \tg{frequency} of cooperation as a function of dilemma strength and reasoning cost with the complete strategy space. Panel \tg{b} shows the \tg{frequency} of cooperation as a function of the dilemma strength and reasoning cost with the limited strategy space.

\noindent\textbf{Figure 3. Cheap talk sustains cooperation through deliberative strategies at low reasoning costs, but its effect vanishes at high costs.} Panel \tg{a} shows the frequencies of each strategy and cooperation as a function of the reasoning cost in the weak dilemma strength regime. Panels \tg{b-d} show the \tg{temporal} evolution of each strategy with different reasoning costs, respectively. The \tg{shaded region indicates} where cooperation dominates. Results are obtained with the parameters $r=0.02$ \tg{throughout}; \tg{panel b uses} $\gamma=0.02$; \tg{panel c uses} $\gamma=0.06$, and \tg{panel d uses} $\gamma=0.15$.

\noindent\textbf{Figure 4. Persistence of cooperation is catalysed by cyclic dominance involving suspicious cooperation (NDC).} \tg{Panels a-e show snapshots of the evolutionary process at time steps 0, 20, 70, 100, and 1000 for $\gamma=0.06$. Panels f-j show snapshots at time steps 0, 15, 70, 100, and 1000 for $\gamma=0.15$.} Results are obtained with parameters $r=0.02$ \tg{throughout}.

\noindent\textbf{Figure 5. A valley of defection induced by moderate reasoning costs.} Panel \tg{a} shows the frequencies of each strategy and cooperation as a function of reasoning cost in strong dilemma strength. Panels \tg{b-d} show the temporal evolution of each strategy with different reasoning costs, respectively. The \tg{shaded region indicates} where cooperation dominates. Results are obtained with the parameters $r=0.2$ \tg{throughout}; \tg{panel b uses} $\gamma=0.02$; \tg{panel c uses} $\gamma=0.1$, and \tg{panel d uses} $\gamma=0.18$.

\noindent\textbf{Figure 6. Strategic defection undermines the persistence of cooperation.} \tg{Panels a-e show snapshots of the evolutionary process at time steps 0, 15, 70, 100, and 1000 for $\gamma=0.1$. Panels f-j show snapshots at the same time steps for $\gamma=0.18$. } Results are obtained with parameters $r=0.2$ \tg{throughout.}

\noindent\textbf{Figure 7. When reasoning cost is small, cooperation is sustained via complex and dynamical co-existence patterns during the \tg{evolutionary} process.} \tg{Panels a and b show snapshots for a weak dilemma with $r=0.02$. Panels c and d show snapshots for a strong dilemma with $r=0.2$.} Results are obtained with parameters $\gamma=0.02$ \tg{throughout.}

\noindent\textbf{Figure 8. \tg{Additional} pathways for cooperation emerge in the complete model.} The parameter regions indicating the strategy with the highest abundance in a square lattice population,  for (a) the complete 8-strategy model and (b) the limited 4-strategy model. The solid lines indicate, $l_1:\gamma=\frac{1}{3}$, $l_2:\gamma=\frac{6r+3}{16}$, $l_3:\gamma=\frac{-6r+7}{16}$, $l_4:\gamma=\frac{8}{21}$, $l_5:\gamma=\frac{6r+3}{32}$, $l_6:\gamma=\frac{-15r+11}{32}$. The complete 8-strategy model contains strategies ACC, ACD, ADC, ADD, NCC, NCD, NDC, and NDD; the limited 4-strategy model contains strategies ACC, ACD, ADD, and NDD.

\bibliographystyle{unsrt}
\bibliography{mybib}

\end{document}